\documentclass[reprint]{revtex4-1}
\usepackage{graphicx}
\usepackage{dcolumn}
\usepackage{bm}
\usepackage{amsmath}
\usepackage{amssymb}

\begin{document}

\preprint{}

\title{Josephson oscillations of two weakly coupled Bose-Einstein condensates}

\author{Dr.~Alexej~Schelle}
\affiliation{Senior Lecturer @ IU Internationale Hochschule, Juri-Gagarin-Ring 152, D-99084 Erfurt}

\date{\today}

\begin{abstract}

A numerical experiment based on a particle number-conserving quantum field theory is performed for two initially independent Bose-Einstein condensates that are coherently coupled at two temperatures. 
The present model illustrates ab initio that the initial phase of each of the two condensates doesn't remain random at the Boltzmann equilibrium, but is distributed around integer multiple values of $2\pi$ from the interference and thermalization of forward and backward propagating matter waves. 
The thermalization inside the atomic vapors can be understood as an intrinsic measurement process that defines a temperature for the two condensates and projects the quantum states to an average wave field with zero (relative) phases.
Following this approach, focus is put on the original thought experiment of Anderson on whether a Josephson current between two initially separated Bose-Einstein condensates occurs in a deterministic way or not, depending on the initial phase distribution.

\begin{description}
\item[Purpose]
Arxiv version of the work.
\end{description}
\end{abstract}

\maketitle

\section{Introduction}

Bose-Einstein condensates have opened the path for the study of very different quantum mechanical phenomena on the micrometer scale, 
building the bridge between microscopic quantum mechanics and classical physics of macroscopic objects \cite{ref-0, ref-1}. 
One major advantage of building experimental setups with Bose-Einstein condensates is the system's controllability and phase coherence on a micrometer scale while minimizing the impact of external noise.
Coherently coupled and tunneling matter waves have e. g. been observed in atomic superconducting crystals as a consequence of coupling two superconductors 
with a thin insulator material, known as the Josephson effect that has been successfully reproduced with Bose-Einstein condensates at temperatures 
that are about ten orders of magnitude colder than standard environmental conditions required for developing superconducting materials \cite{ref-2}.
It is the collectively coherent character of ultra-cold atomic condensed matter that has in particular enabled the measurement of relative phase and particle imbalances 
of atomic quantum states and wave functions, respectively, with interfering partial matter waves built from Bose-Einstein condensates \cite{ref-3, ref-4, ref-5}, also known as atomic lasers \cite{ref-6}.

Motivated to understand the concept of broken gauge symmetry in solids, Anderson and later on Leggett built an interesting thought experiment on the role of the absolute phase 
of quantum states to the phenomenon of spontaneous gauge symmetry breaking below a critical temperature \cite{ref-7, ref-8}. 
Spontaneous symmetry breaking is nowadays understood as an effect of disordering existing global symmetries in a quantum system to asymmetric configurations which do 
not allow global gauge phase symmetry anymore after the system has passed a critical point and formed a new aggregate state, such as the transition from fluid to gaseous phases 
in water.
Phase transitions are described mainly with order parameters that quantitatively indicate the transition from one aggregate phase to another in that the value of the order parameter 
rapidly changes as a function of system-specific parameters such as temperature \cite{ref-9, ref-10}.
To understand the questions raised by the famous thought experiment of Anderson, let's recall and think about what happens if we bring two separated and independently formed Bose-Einstein condensates (at different temperatures)
in contact by placing a weak link between the two trapping potentials. 
In such a case, the absolute phase of the two objects may become meaningful considering a measurement process that measures the relative phase between the condensates \cite{ref-5}. 
However, the outcome of a measurement for the absolute as well as the relative phase is not standard in formulations of quantum mechanics, since depending on the random phases of the Bose-Einstein condensates, a different Josephson current may flow as a function of the relative phase between the two atomic ensembles. 
This is because according to the theory of quantum mechanics, if we assume the absolute phase of the two condensates to be well-defined, but random, a Josephson current will flow at any trial, however, the initial phase and thus the initial strength of the current will remain random. 

In contrast, real experiments on interfering Bose-Einstein condensates that measure the relative phase of the two condensates indicate that the initial relative phase that leads to Josephson oscillations 
is not random, but distributed around values of zero phase \cite{ref-11} when lowering the potential barrier of the initially completely separated Bose-Einstein condensates.
While standard theoretical approaches can only explain the occurrence of a well-defined phase from the measurement process itself, one has to assume a certain initial value for the relative phase to calculate the resulting Josephson oscillations.
Hence, so far theoretical approaches cannot derive the value of relative initial phases associated with deterministic Josephson currents ab initio for two weakly coupled Bose-Einstein condensates within standard quantum-theoretical models. 
Is the initial absolute phase possibly pre-defined from an internal measurement process before external measurements of the initial phase take place?

The concept of spontaneous symmetry breaking for Bose-Einstein condensates indicates that the assignment of an absolute phase (with zero average) of the total field can only take place by the thermalization of the random quantum field to the Boltzmann equilibrium \cite{ref-10}.
Because, as shown before, symmetry breaking (phase separation) of the quantum field occurs mainly in the relation between condensate and non-condensate quantum field components of the Bose-Einstein condensate, and the global phase gauge symmetry of the quantum field remains preserved.
Within the present model, it is further illustrated that the initial relative phases can be numerically calculated ab initio within a number-conserving quantum field theory that accurately models the conservation of the particle number and the coherence between forward and backward propagating quantum fields at finite temperature \cite{ref-6, ref-13}.
It is shown numerically that the initial phase distribution for the equation of motion that describes the dynamics of Josephson's weak link is not random, but distributed around multiples of the circle number $2\pi$ - while the global gauge symmetry of the total quantum field remains preserved. 
Numerical results are obtained within Monte-Carlo simulations for the correlation and relative phase distribution. 
Quantization of the initial phase arises without further theoretical assumptions on the time variable from the quantum field model, indicating the interference of partial wave fields between the two components in the double-well potential. 

The presented quantum field model in particular formally defines an intrinsically and numerically derived time scale (coherence time) for which disjoint forward and backward propagating wave fields are equal.  
At the thermal Boltzmann equilibrium that builds the foundations for a definite zero average absolute phase of the quantum field, the two counter-propagating wave fields interfere (are equal) at multiples of the oscillation period of the quantum fields in the complex plane. 
The latter fact is mathematically accounted for by projecting the non-condensate quantum states onto the Boltzmann equilibrium \cite{ref-14} which can be interpreted as an internal measurement process that measures the temperature of the atomic cloud realized by the rapid thermalization 
within each of the considered components as assumed within the theoretical model. 
The numerical technique has recently been also applied for coherent atom lasers built from single-component ultra-cold atomic matterwaves confined in an external harmonic trapping potential \cite{ref-6} and confirms the experimental findings. 

\section{Theory}

In the following, two locally separated and independently created Bose-Einstein condensates as presented in Ref.~\cite{ref-11} are considered. 
Much the same as in the real experimental setup, we assume that the two Bose-Einstein condensates can be prepared at defined temperatures and slowly be brought into 
contact by lowering the potential barrier such as to realize a Josephson weak link between two initially separated Bose-Einstein condensates of ideally zero initial particle number imbalance.
According to the considerations and thought experiment in the introduction, if one applies the principles of standard quantum mechanics, each of the two Bose-Einstein condensates 
should in principle be associated with one global and random phase related to the wave function that describes the local distribution of the wave field created by the ensembles of atoms.
However, if this is the case, it is not clear why the relative phase is distributed only around values of zero phase, as observed in the real experiments on interfering Bose-Einstein condensates interacting with a weak Josephson link.

To answer this question, it is possible to model the quantum fields of the two Bose-Einstein condensates as derived in the framework of the number-conserving quantum field theory described in Refs.~\cite{ref-6, ref-10}.       
Taking the spatial average over local quantum fields leads to the non-local order parameters

\begin{equation}
\label{eq.1}
\chi_{1,2}(t) = \sum_{\bf{k}}c^{(1,2)}_{\bf{k}}{\rm e}^{-i\phi^{(1,2)}_{\bf{k}}(t)}
\end{equation}\\
that describes the quantum field's time evolution in the two potential barriers.
Since thermal equilibrium is assumed for the calculation of the quantum fields and related properties (not for the chemical potentials associated with the phases that arise in Eq. (\ref{eq.1})), the explicit time dependence can be changed by the temperature-dependent parameter $\beta$, 
and numerical quantities of the model are defined up to a specific uncertainty measure. 
Please note that simply assuming time propagation of the wave functions leads to the decay of the associated coherent wave fields. 
Instead, in the derivation of the equation for the quantum fields, it is assumed that the Bose-Einstein condensate approaches a thermal Boltzmann equilibrium after each interaction process of particles within the Bose gas, also below the critical temperature.
Consistently, the quantum state of the wave field is projected onto thermal equilibrium using the concept of complex time $it = \hbar\beta$.

Similar to the formal description of atom lasers, trap parameters of the microwave cavity are characterized by trap frequencies $(\omega^{(1,2)}_x, \omega^{(1,2)}_y, \omega^{(1,2)}_z)$ and single particle energies $\epsilon^{(1,2)}_{\bf{k}} = \hbar(k^{(1,2)}_x\omega^{(1,2)}_x + k^{(1,2)}_y\omega^{(1,2)}_y + k^{(1,2)}_z\omega^{(1,2)}_z)$. 
In the present setup, instead of modeling a resonantly external driving force that acts onto the total quantum field, we model the coherent coupling of two \textit{different} Bose-Einstein condensate components (that can be experimentally realized e. g. with an ensemble of separated clouds of $^{87}{\rm Rb}$ atoms with an s-wave scattering 
length of $a = 5.4 {\rm~nm}$ Bose-Einstein condensate in a double-well potential).
To describe correlations of the two initially separated Bose-Einstein condensates that may e. g. arise from Josephson weak links, it is further numerically implemented that the wave fields are coherently coupled, i. e. lowering the potential barrier (or inducing coherent excitations in spinor Bose-Einstein condensates)
may result in two coupled wave fields that are mathematically well-described in the form

\begin{equation}
\label{eq.2}
\chi(\beta, \Delta\phi) = \chi_{1}(\beta) + {\rm e}^{i\Delta\phi}\chi_{2}(\beta) \ ,
\end{equation}\\
where $\chi_{1}(\beta)$ and $\chi_{2}(\beta)$ are the two non-local (spatially averaged) quantum fields represented by complex numbers in the Gaussian number plane 
and $\Delta\phi$ is the relative phase between the two macroscopic wave functions of the two atomic vapors.

To quantify correlations numerically, we further define the correlation function

\begin{equation}
\label{eq.3}
c(\chi^\star, \chi) = \chi^*(\Delta\phi, \beta)\chi(\Delta\phi, \beta)
\end{equation}\\
that describes correlations between the two quantum fields as a function of the relative phase $\Delta\phi$. 
To account for particle number conservation, the equality 

\begin{equation}
\begin{split}
\label{eq.4}
&\sum^{\infty}_{j=0} z^{j}(\mu^{(1,2)})\left[\prod_{l=x, y, z} \frac{1}{1-{\rm e}^{-j\beta^{(1,2)}\hbar\omega_l^{(1,2)}}} - 1\right] = 
\\\\ &=  N^{(1,2)} - N^{(1,2)}_0
\end{split}
\end{equation}\\
is implemented numerically, with $z(\mu^{(1,2)}) = {\rm e}^{\beta^{(1,2)}\hbar\mu^{(1,2)}}$ the fugacity of the corresponding Bose-Einstein condensates in the potential wells 1 and 2, respectively, 
$N^{(1,2)}$ the total atom number, $N^{(1,2)}_0$ the condensate atom number, $\beta^{(1,2)}$ the inverse thermal energy and $\omega_{\bf{k}}^{(1,2)}$ the photon frequency of the external confinement (in mode direction $x, y, z$) of either one of the two Bose-Einstein condensates.
The mathematical phases in Eq.~(\ref{eq.1}) can be directly related to the solutions of Eq.~(\ref{eq.4}), i. e. to the chemical potentials $\mu^{(1,2)}_{\bf{k}}$ of the atomic quantum field modes, through the relation $\phi^{(1,2)}_{\bf{k}} = {\rm Re}\lbrace\frac{-\beta\mu^{(1,2)}_{\bf{k}}}{\hbar}\rbrace$. 

Numerical calculation of correlations $c(\Delta\phi,\beta)$ in Eq. (\ref{eq.3}) are obtained from sampling the complex-valued quantum fields according to the equation

\begin{equation}
\label{eq.5}
\chi^{(1,2)} = \vert\psi\vert{\rm e}^{i\phi^{(1,2)}} = \sum_{\bf{k}}c^{(1,2)}_{\bf{k}}{\rm e}^{-\beta^{(1,2)}\mu^{(1,2)}_{\bf{k}}} \ ,
\end{equation}\\
where $\lbrace c^{(1,2)}_{\bf{k}}\rbrace$ are random complex valued weighting factors.
Relative weighting of the different quantum field realizations is obtained from Monte-Carlo sampling with weighted partial probability amplitudes according to the probability distributions

\begin{equation}
\label{eq.6}
p^{(1,2)}(\mu) =  \frac{{\rm e}^{-\beta^{(1,2)}\vert\mu^{(1,2)}\vert}}{\mathcal{N}^{(1,2)}_{\mu^{(1,2)}}} \ 
\end{equation}\\
with $\mathcal{N}^{(1,2)}_{\mu^{(1,2)}} = \int_{\mathbb{C}}{\rm d}\mu^{(1,2)}~{\rm e}^{-\beta^{(1,2)}\vert\mu^{(1,2)}\vert}$.
The total chemical potentials of the two different Bose-Einstein condensates are numerically calculated from the average of the partial chemical potentials at defined weighting amplitudes $\lbrace c^{(1,2)}_{\bf{k}}\rbrace$ for each realization of the Monte-Carlo sampling algorithm.
Additionally, random sampling of the relative initial phase $\Delta\phi$ between the two interacting components allows the analysis of the quantitative scaling for the correlation function in Eq. (\ref{eq.3}), as a function of the initial relative phase.
At this stage, no assumptions are made on the initial phase distribution for the different sampling steps.
Hence, correlations are numerically quantified ab initio from the interference of the two locally averaged quantum fields induced by the distribution of the different coherent phases (partial chemical potentials) that depend on the shape of the external confinements and the temperatures of the atomic vapors. 

Calculating the absolute value of the correlation function in Eq. (\ref{eq.3}) defines a real-valued probability distribution $\Pi(\Delta \phi)$ from the complex-valued phase correlation function,

\begin{equation}
\label{eq.7}
\Pi(\Delta \phi) = \vert c (\Delta \phi) \vert = \sqrt{{\rm Re}^2(\chi^\star\chi) + {\rm Im}^2(\chi^\star\chi)}
\end{equation}\\
with ${\rm Re}(\chi^\star\chi)$ the real part and ${\rm Im}(\chi^\star\chi)$ the imaginary part of the correlation function $c(\chi^\star,\chi)$.

\section{Numerical simulations}

\begin{figure}[t]
\begin{center}
\includegraphics[width=7.0cm, height=5.0cm,angle=0.0]{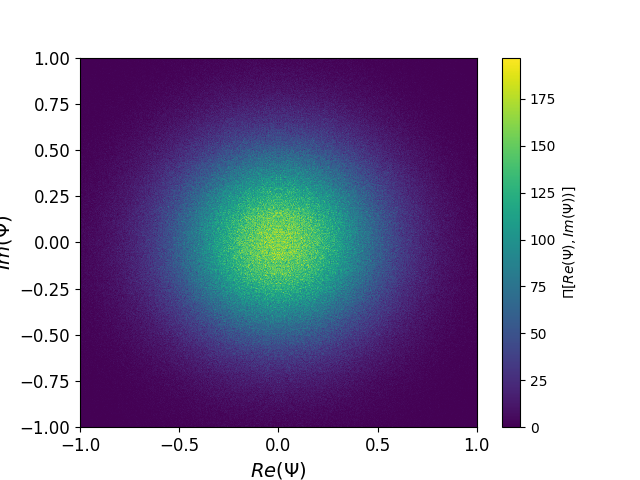} \\
\caption{(Color online) Figure highlights realizations of a local average quantum field in complex number representation for a Bose-Einstein condensate confined in an 
external trapping potential with trap parameters $\omega^{(1,2)}_x = 2\pi\times250 {\rm~Hz}$, $\omega^{(1,2)}_y = 2\pi\times150 {\rm~Hz}$ und $\omega^{(1,2)}_z = 2\pi\times50 {\rm~Hz}$ at temperature $T = 10$ nK obtained from 
a Monte-Carlo simulation of Eq. (\ref{eq.1}).}
\label{fig.1}
\end{center}
\end{figure} 

Numerical simulations for quantum phase correlations as defined in Eq. (\ref{eq.3}) are performed for parameters that describe the quantum field distributions and correlations, respectively, as a function of temperatures $T^{(1,2)}$ 
with trap parameters $\omega^{(1,2)}_x$, $\omega^{(1,2)}_y$ and $\omega^{(1,2)}_z$ at given total particle numbers $N^{(1,2)}$ and condensate particle number $N^{(1,2)}_0$. 
The quantum field distribution for sampling realizations of the locally averaged quantum field of a Bose-Einstein condensate in complex number representation as defined in Eq. (\ref{eq.1}) 
can be compared to the quantum field which represents the sum of two coherent quantum fields that are correlated by numerically modeling uniformly randomized relative phases between the two quantum fields.
The coherent coupling of two quantum fields can be e. g. be realized experimentally by a coherent laser source that resonantly couples the energetic transition for the atoms in the initially locally separated Bose-Einstein condensates (which are spatially localized in the two minima of a double-well potential). 
To model the quantum field $\chi(\beta)$ as shown in Fig. \ref{fig.1} for a Bose-Einstein condensate below the critical temperature that defines the distribution of the field $\Pi[{\rm Re}(\chi), {\rm Im}(\chi)]$, an
external trapping potential with trap parameters $\omega_x = 2\pi\times250 {\rm~Hz}$, $\omega_y = 2\pi\times150 {\rm~Hz}$ und $\omega_z = 2\pi\times50 {\rm~Hz}$ at temperature $T = 10$ nK is assumed 
for Monte-Carlo sampling of the field.  

\begin{figure}[b]
\begin{center}
\includegraphics[width=7.0cm, height=5.0cm,angle=0.0]{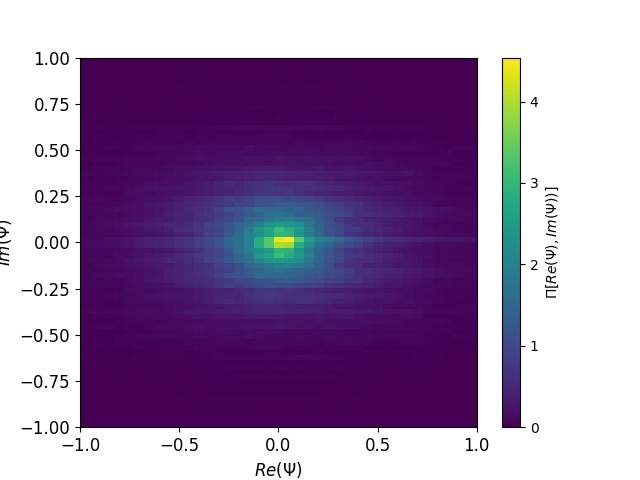} \\
\caption{(Color online) Figure highlights interference effects with reduced volume in the generalized phase space (${\rm Re}(\chi), {\rm Im}(\chi)$). 
Numerical realizations of local average quantum fields are obtained from Eq. (\ref{eq.2}) for two coherently coupled Bose-Einstein condensates in complex number representation confined in 
two different external trapping potentials with the same trap parameters as in Fig. \ref{fig.1} at temperatures $T_1 = 10$ nK (left well) and $T_2 = 125$ nK (right well).}
\label{fig.2}
\end{center}
\end{figure}

\begin{figure}[t]
\begin{center}
\includegraphics[width=4.0cm, height=3.0cm,angle=0.0]{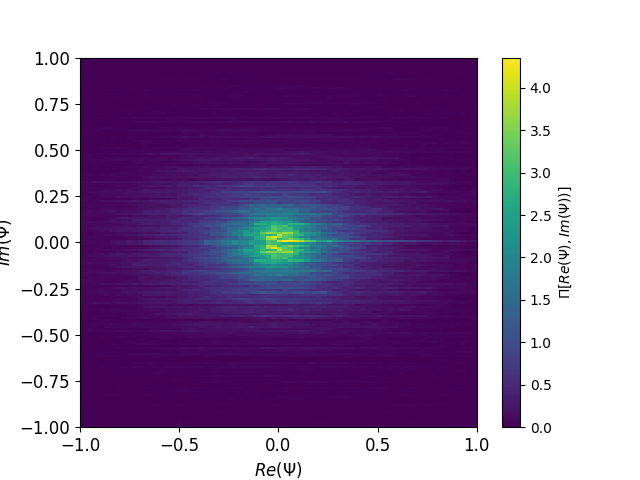} 
\includegraphics[width=4.0cm, height=3.0cm,angle=0.0]{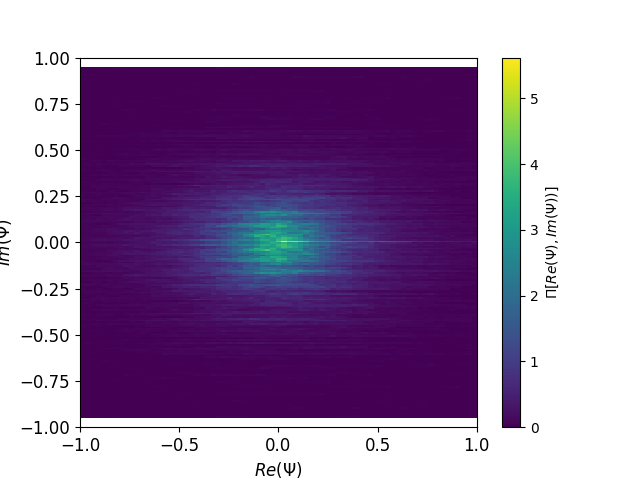} \\
\includegraphics[width=4.0cm, height=3.0cm,angle=0.0]{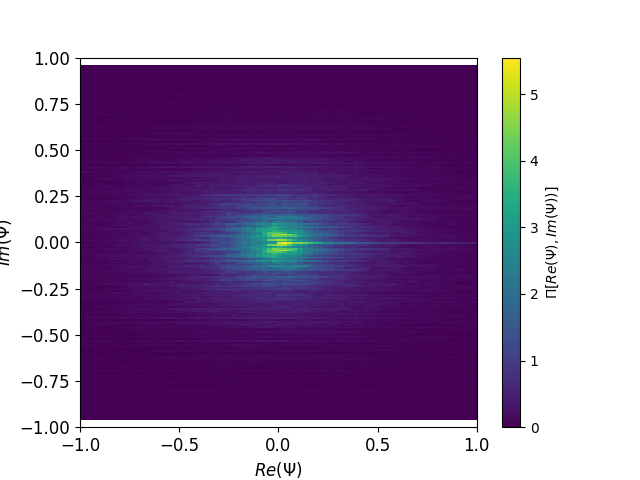} 
\includegraphics[width=4.0cm, height=3.0cm,angle=0.0]{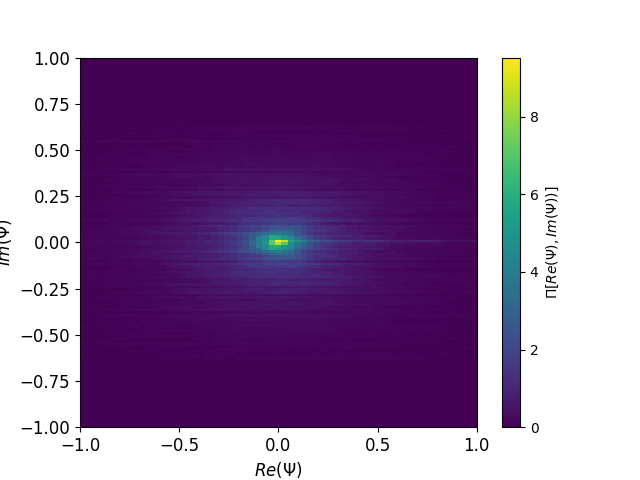} \\
\caption{(Color online) Shown are random fluctuations of the quantum field for different sample realizations of two coupled Bose-Einstein condensates at the same parameters as in Fig. \ref{fig.1}.
As indicated by the simulations, the generalized phase space volume, as well as the intensity of the two interfering wave fields, varies in each sample realization, i.e. 
as a function of the initial conditions for the quantum field.}
\label{fig.2a}
\end{center}
\end{figure}

To highlight the effects of matter-wave interference due to the difference in coherence times (oscillation periods of the quantum fields in the complex plane) as a function of temperature, 
coherently coupled quantum fields as defined by Eq. (\ref{eq.2}) are modeled at two different temperatures $T = 10$ nK (left well) and $T = 125$ nK (right well).   
Calculating the interference of the two numerically so-obtained and coherently coupled quantum fields leads to a quantum field distribution that is reduced in volume in the generalized phase space representation $[{\rm Re}(\chi), {\rm Im}(\chi)]$ by several orders of magnitude as shown in Figs. \ref{fig.2} and \ref{fig.2a}.
This results from destructive interference of different field modes that are weighted with random interfering field phases added to the field amplitudes in Eq. (\ref{eq.5}) for components of the quantum field with larger absolute values of the quantum field (outer circle of the quantum field) and from 
constructive interference of the different field modes that are distributed close to the zero point (inner circle of the quantum field).
Randomization of the relative phases thus mainly leads to destructive interference at the edges of the quantum field distribution (out of the Boltzmann equilibrium), where the absolute value of the imaginary part of the chemical potentials that define the relative phases in the absence of external coupling is 
zero.
Quantum field states with larger values of the chemical potential's imaginary part (corresponds to $\vert z\vert = 1$ in the representation of the associated fugacity in the complex plane, compare \cite{ref-10}), 
lead to constructive interference and non-vanishing numerical values of the composite quantum field (i. e. the direct sum of condensate and non-condensate field components). 
The intensity and volume of the generalized phase space regions change with the temperature and trap geometry of the two atomic clouds.

Analyzing the quantum field distributions concerning the interference of two coupled quantum fields numerically again draws back attention to Anderson's original thought experiment that asks why the interference of 
two initially separated Bose-Einstein condensates do not lead to completely destructive interference on average as expected, since both phases of the two Bose-Einstein condensates 
are in principle completely random?
Shouldn't the random offset values of the two initial phases also randomize the total relative phase between the two condensates, and as a consequence, a current with random field strength and orientation should flow at any trail?

\begin{figure}[b]
\begin{center}
\includegraphics[width=4.0cm, height=3.25cm,angle=0.0]{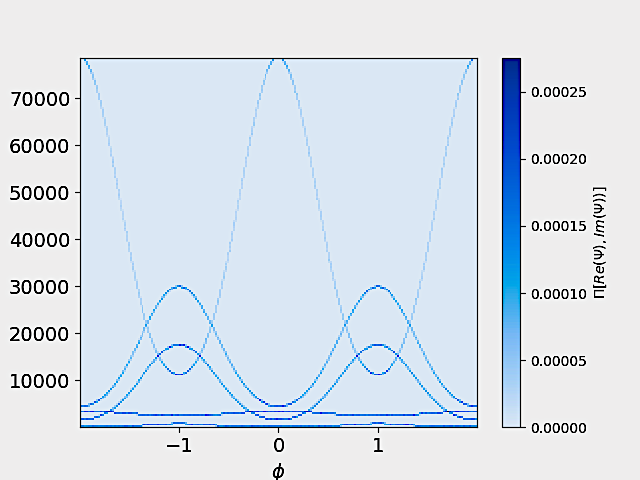} 
\includegraphics[width=4.0cm, height=3.25cm,angle=0.0]{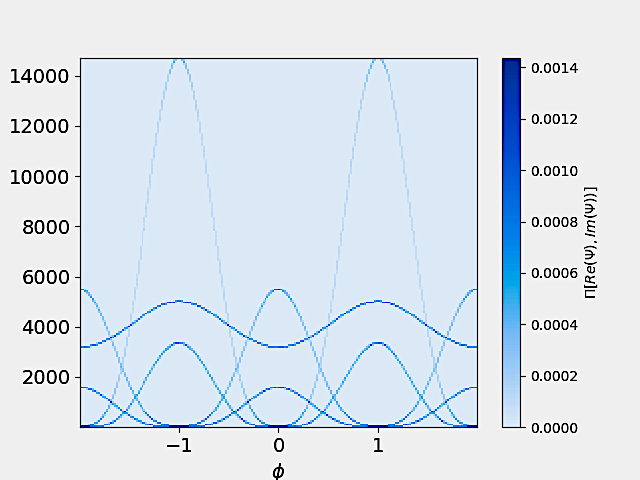} \\
\includegraphics[width=4.0cm, height=3.25cm,angle=0.0]{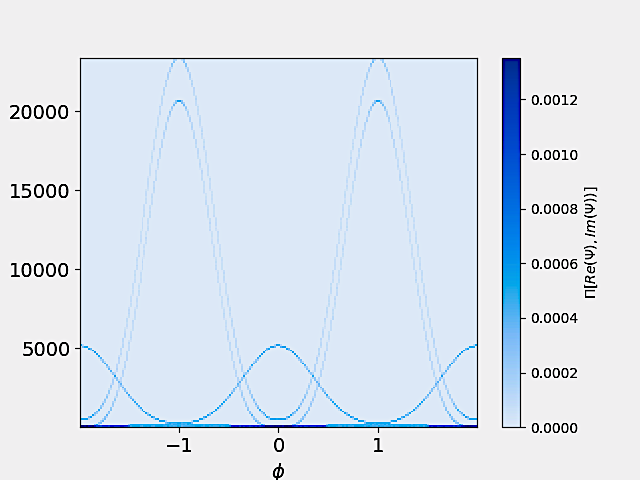} 
\includegraphics[width=4.0cm, height=3.25cm,angle=0.0]{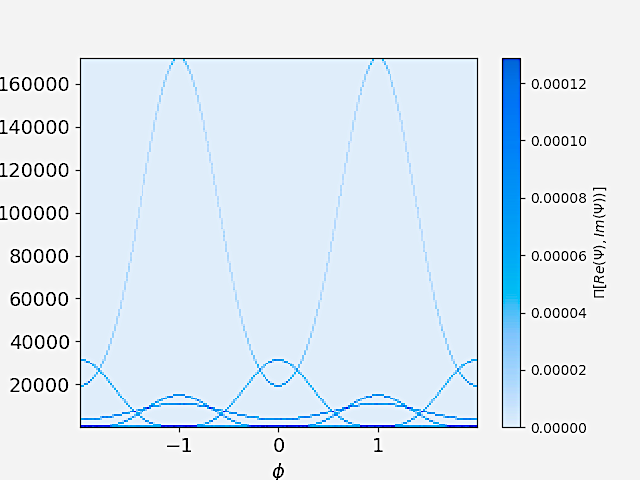} 
\caption{(Color online) Figure highlights different numerical model realizations of the correlation function in Eq. (\ref{eq.3}) for
always the same parameters of Fig. \ref{fig.1} with temperatures $T_1 = 10$ nK (left well) and $T_2 = 25$ nK (right well). 
As the results indicate, the initial phase $\Delta \phi_0$ is always distributed around multiples of the circle number $2\pi$.
This illustrates non-deterministic correlations with deterministic and well-defined initial phase distributions around $\Delta \phi_0 = k\times2\pi$ from 
constructive and destructive interference of partial matter waves that follow complex spectra $\Omega^{(1,2)}(\mu^{(1,2)}) = \Omega^{(1,2)}(\omega) + \Omega^{(1,2)}(\Gamma) = \lbrace\omega^{(1,2)}_{\bf{k}} + i\Gamma^{(1,2)}_{\bf{k}} \rbrace$.
Small offset values of some realizations of the correlation function indicate that the phase is weakly randomized by non-deterministic background fluctuations (from temperature).
The scaling behavior is universal for different temperatures and trap geometries.}
\label{fig.3}
\end{center}
\end{figure}

Within the present model, the internal thermalization process projects the quantum state of the Bose-Einstein condensate to the thermal Boltzmann equilibrium that corresponds to quantum field states in the vicinity of ${\rm Re}(\chi) = 1$ and ${\rm Im}(\chi) = 0$. 
In the following, four ensembles of different realizations are further shown, as defined numerically exactly at the same initial conditions and external trap parameters, compare Fig. \ref{fig.3} for 
the same parameters of Fig. \ref{fig.1} at any realization given the temperatures $T=10$ nK (left well) and $T= 25$ nK (right well).
Considering numerically the quantum field correlation function in Eq. (\ref{eq.3}) as a function of uniformly randomized relative initial phases thus indicates that correlations of the two coupled quantum field components in the representation of the field in the complex plane 
are distributed around multiples of the circle number $2\pi$, as indicated in Fig. \ref{fig.3}.
In particular, as illustrated in Fig. \ref{eq.2}, the total average of the two interacting quantum fields is non-zero at the center of the distribution.     
Small offset values of some realizations of the correlation function indicate that the relative initial phase is (only) weakly randomized by non-deterministic background fluctuations resulting from the atomic vapors' temperatures.
Since the realization of the quantum fields differs from each ensemble of field realizations, the distribution of the strength of the correlation between the two quantum fields is non-deterministic, i. e. always different in shape, however, 
the average relative initial phase difference is well-defined and predictable.

As a function of time, after preparation of an initial state according to the quantum field at thermal equilibrium obtained from Monte-Carlo simulations, the out-of-equilibrium dynamics for 
two weakly coupled Bose-Einstein condensates in a double-well potential can be described in terms of the standard equation of motion for the relative phase and particle number imbalance, 
given the initial values for the relative phase as a boundary condition.
Applying the non-local quantum field representation of the present field theory, up to the first order in $\Delta z$, the relative phase of the two condensates follows the relation

\begin{equation}
\label{eq.8}
\frac{{\rm d}\phi}{dt} = \Omega\Delta z + 2J\Delta z\times{\rm cos~}\phi \ ,
\end{equation}\\
where $\Omega$ is the energy offset (in units of the Planck constant $\hbar$) between the left and right well, $J$ the dimensionless coupling strength between the two condensates, $t$ defines the time variable of the quantum system, $\Delta z$ the particle number imbalance and $\Delta \phi_0$ is the initial phase difference between the two atomic vapors.
At the initial time $t=0$ for the measurement of the relative initial phase $\Delta\phi_0$ and particle number imbalance $\Delta z$, zero particle number imbalance is assumed in the present study to draw the effects of phase correlations between 
the two quantum fields numerically represented by Monte-Carlo simulations.  
The corresponding (coupled) equation of motion for the particle number imbalance $\Delta z$ reads

\begin{figure}[t]
\begin{center}
\includegraphics[width=6.0cm, height=4.5cm,angle=0.0]{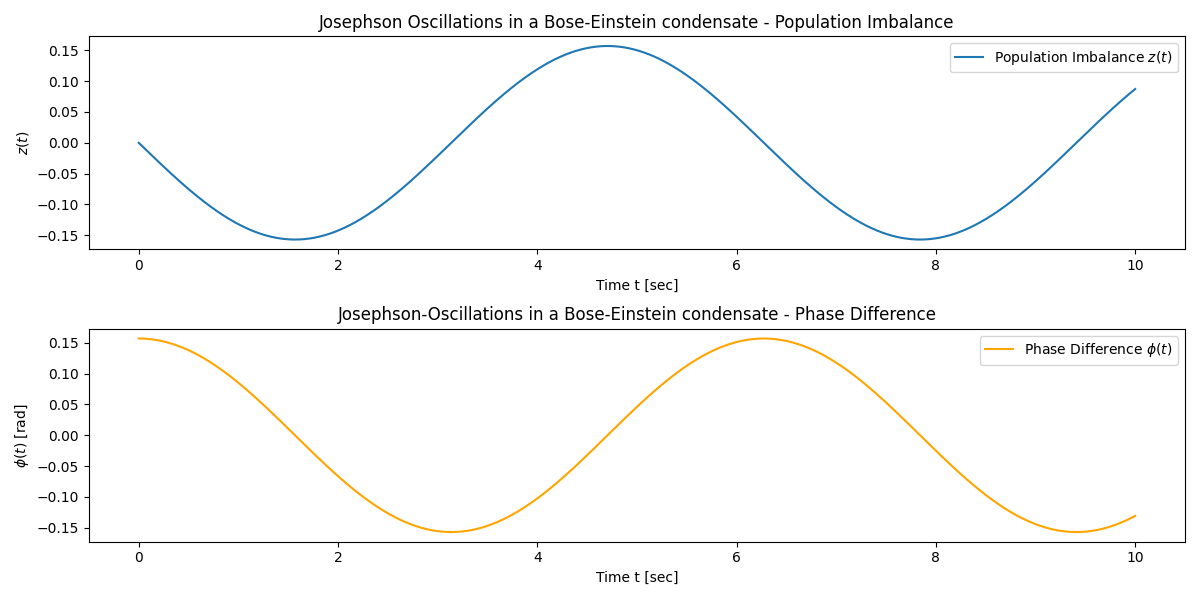} \\
\includegraphics[width=6.0cm, height=4.5cm,angle=0.0]{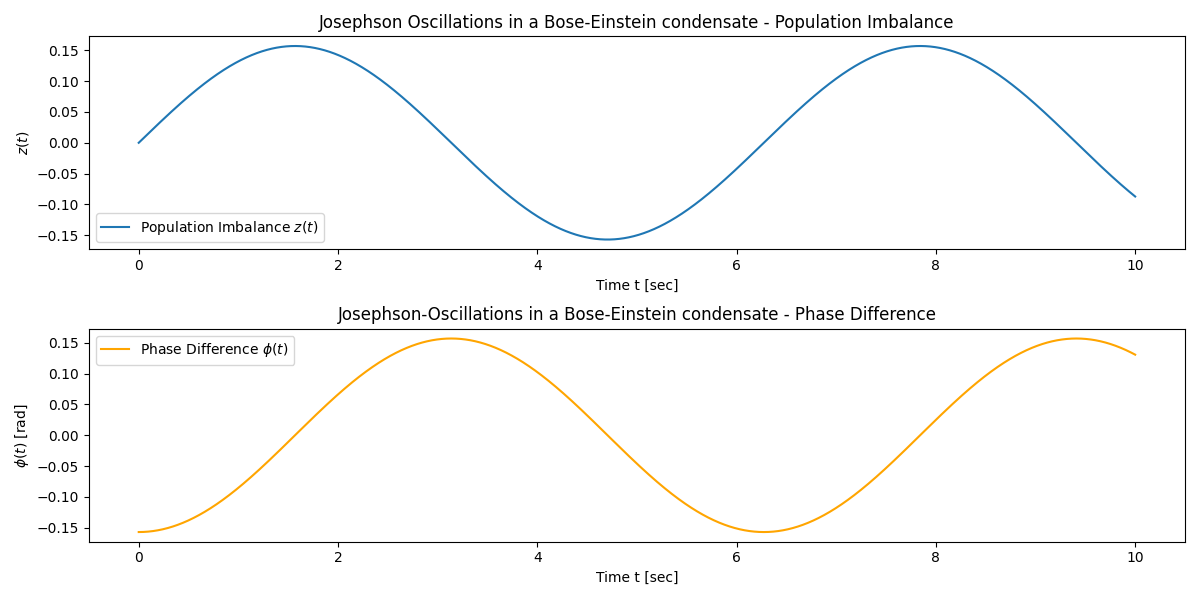} \\
\caption{(Color online) Shown are Josephson oscillations as a function of initial particle imbalance (orange line) and phase difference (blue line) for two weakly coupled Bose-Einstein condensates in a double-well potential. 
Parameters are set to coupling strength of $J = 0.5$ and initial particle number imbalance $\Delta z = 0$ with initial phase difference $\Delta \phi_0 = 0.05$ (upper figures) and $\Delta \phi_0 = -0.05$ (lower figures). 
Hence, for an initial quantum field realization of zero particle number imbalance and arbitrary small initial phase around the value of $\Delta\phi_0 = 0$ at ideally the same trap geometry, the oscillating phase distribution is shifted by a positive quarter oscillations period ($+\tau/2$), and a Joesphon current 
flows at any trial.}
\label{fig.4}
\end{center}
\end{figure} 

\begin{figure}[t]
\begin{center}
\includegraphics[width=6.0cm, height=4.5cm,angle=0.0]{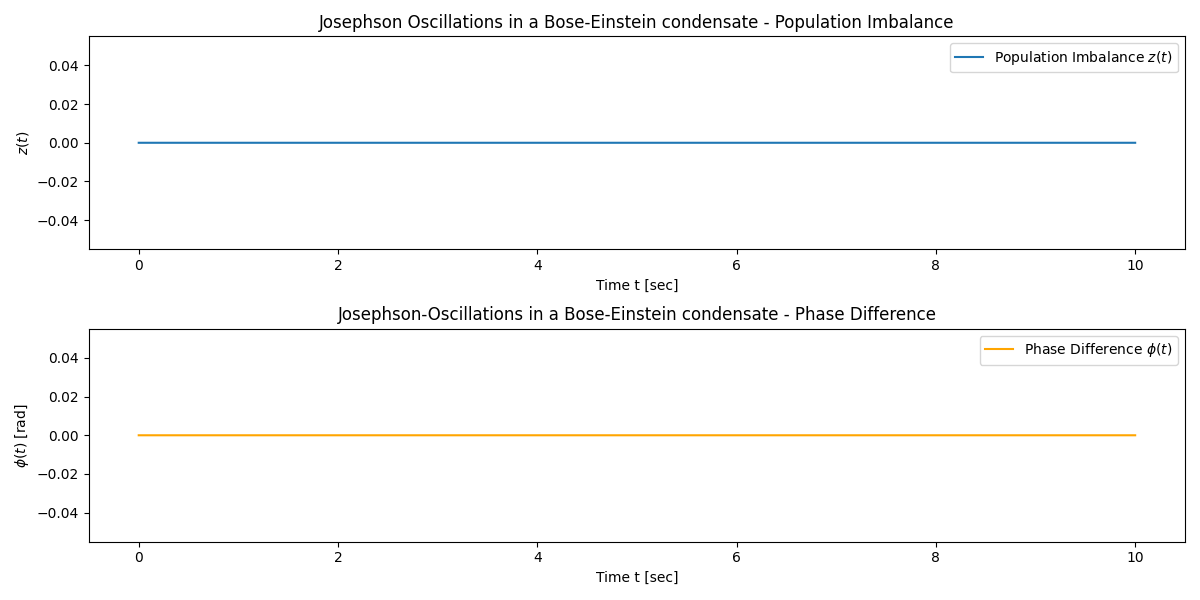} \\
\caption{(Color online) In case of initial zero particle number imbalance at ideally equal trap geometries and zero temperature, there exists precisely one (unstable) configuration with $\Delta \phi_0 = 0$ for which no Joesphon current flows. 
Since the number of possible configurations with non-zero Josephson current (non-zero initial phase difference - compare Fig. \ref{fig.3}) is over-countable, the probability for the sketched configuration effectively tends to zero.
From the Heisenberg principle, it thus follows that a Josephson current also flows for exactly zero relative initial phases ($\Delta\phi_0 = 0$), since it is realistically impossible to achieve a configuration with $\Delta \phi_0 = 0$ and $\Omega = 0$ on the uncertainty scale provided.}
\label{fig.5}
\end{center}
\end{figure}
\begin{figure}[t]
\begin{center}
\includegraphics[width=6.0cm, height=4.5cm,angle=0.0]{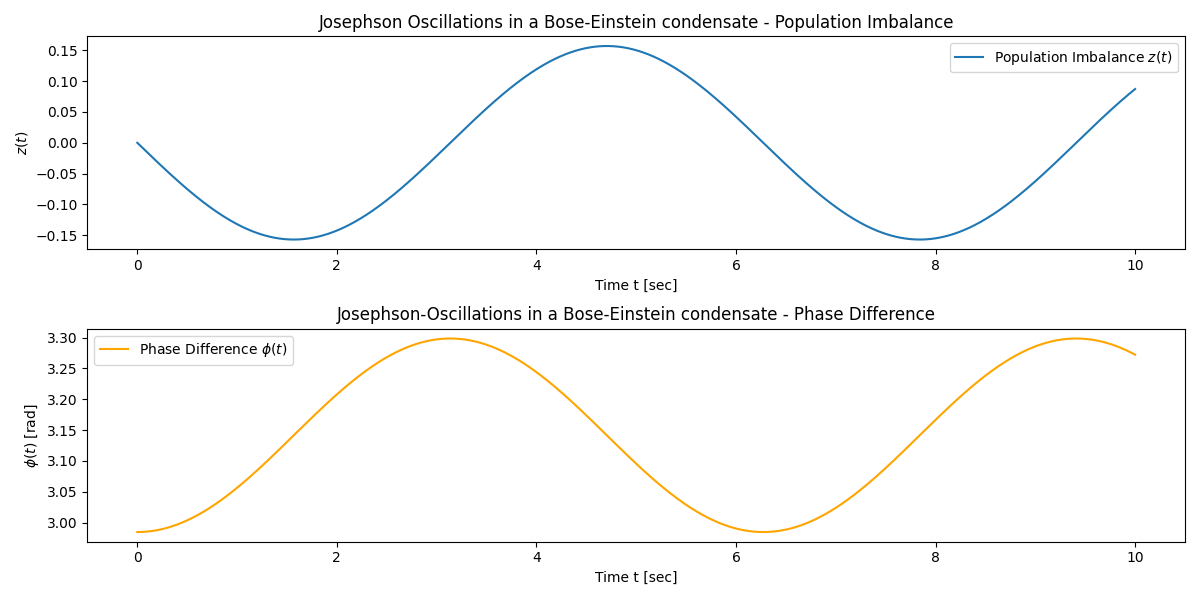} \\
\includegraphics[width=6.0cm, height=4.5cm,angle=0.0]{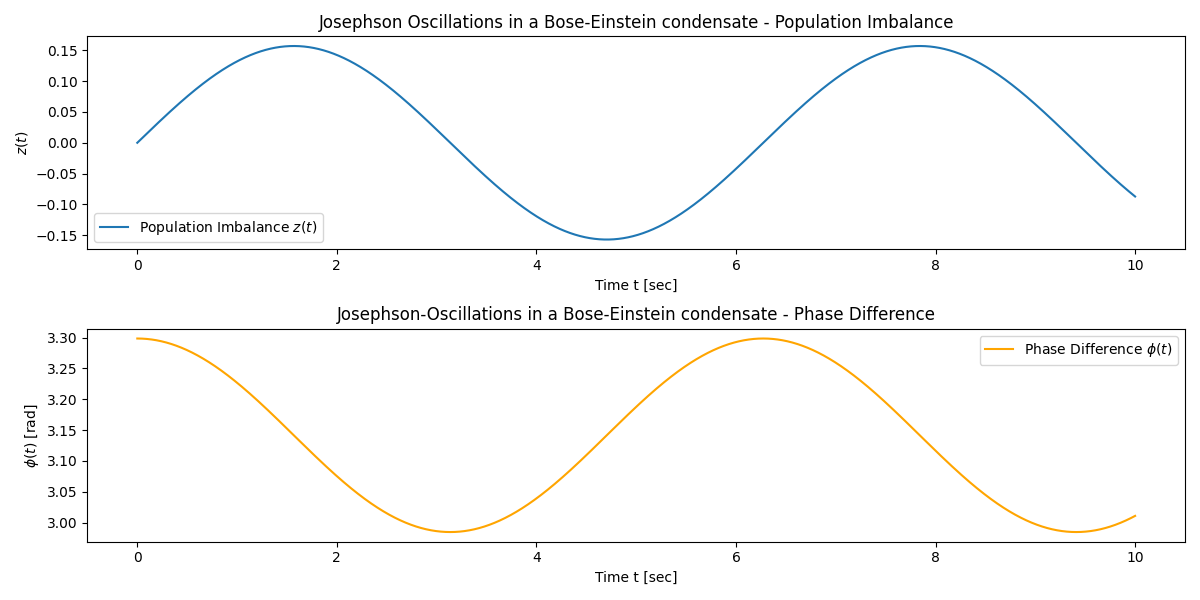} 
\caption{(Color online) Shown are Josephson oscillations (for ideally undamped oscillations) as a function of initial particle number imbalance (orange line) and phase difference (blue line) for two weakly coupled Bose-Einstein condensates in a double-well potential. 
Parameters are set to coupling strength of $J = 0.5$ and initial particle number imbalance $\Delta z = 0$ with initial phase difference $\Delta \phi_0 = 0.95$ (upper figures) and $\Delta \phi_0 = 1.05$ (lower figures). 
Hence, for an initial quantum field realization of zero particle number imbalance and arbitrary small initial phase variations around the value of $\Delta\phi_0 = \pm2\pi$ at ideally the same trap geometry, the oscillating phase distribution is shifted 
by a negative quarter oscillations period ($-\tau/2$), and a Josephson current flows at any trial.}
\label{fig.6}
\end{center}
\end{figure}

 

\begin{equation}
\label{eq.9}
\frac{{\rm d}\Delta z}{dt} = - 2J\times{\rm sin~}\phi \ .
\end{equation}\\
Hence, from the phase correlations highlighted in Fig. \ref{fig.3}, conclusions about Josephson oscillations can be drawn from modeling the initial phase correlations that define the initial conditions for the phase and particle number imbalance.
As illustrated in Fig. \ref{fig.4} and Fig. \ref{fig.6}, for initial conditions with initial particle number imbalance $\Delta z = 0$ and initial phase difference $\Delta \phi_0 = \pm\epsilon$ and $\Delta \phi_0 = \mp\epsilon$ with $\epsilon \rightarrow 0^+$ from the numerical simulations it is expected and confirmed, 
respectively, that a Josephon current with a well-defined initial phase difference of $\epsilon = \pm0.05\pi$ flows between the two different double-well potentials.
Please note that in the case of initial phase difference $\Delta \phi_0 = +\epsilon$, the current flows in the opposite direction as compared to the case, where $\Delta \phi_0 = -\epsilon$.
Depending on the initial phase difference, the oscillating phase is thus shifted by a quarter oscillation period ($\pm\tau/2$).
Starting at the initial phase difference of $\Delta \phi_0 = \pi\pm\epsilon$ the system behaves equivalently, however, the oscillating phase is shifted by a negative quarter oscillation period ($\mp\tau/2$).

\section{Discussion}

From the present model, it is thus possible to predict the correlations and hence distribution of the relative initial phase between two coherently coupled Bose-Einstein condensates.
Coherent coupling may e. g. be realized by experimentally creating double-well trapping potentials with variable heights of the central potential barrier to model a Josephson 
weak link between two or more Bose-Einstein condensates trapped in the minima of the double-well potential.
The analysis of such an experimental setup does ad hoc not provide a clear answer to the question of what precisely defines the initial relative phase difference between the Bose-Einstein condensates.
As shown in the sequel of the present quantitative model for Bose-Einstein condensation, the number-conserving quantum field theory for Bose-Einstein condensation provides a definite prediction 
for the correlations of the relative phase of two coupled Bose-Einstein condensates.
The superposition of coherently coupled atoms in an ultracold atomic vapor can e. g. also be experimentally realized from resonantly driven transitions of different angular momentum 
quantum states that prepare different spinor quantum states with a well-defined initial phase difference from the external driving laser modulus multiples of $2\pi$ for the relative initial phases as discussed.    

The case that no Josephson current is observed after connecting two Bose-Einstein condensates with a Josephson weak link as shown in Fig. \ref{fig.5} can only be observed for the case that the initial particle number imbalance, as well as initial phase difference, is zero given that the trap geometry and the temperatures of the atomic clouds are exactly equal.
The standard theory of quantum mechanics thus basically excludes this situation from the Heisenberg principle, since the parameters of the external confinements are only defined up to a certain uncertainty measure, which means that the possibility of modeling equal external conditions for the two Bose-Einstein condensates is zero 
at the scale defined by the Heisenberg principle. 

Starting from the fundamental principles of the thought experiment motivated by Anderson and Leggett, nothing prevents applying the same numerical calculus of the present model to the initial relative phase correlations 
between forward and backward propagating field components (for each independent quantum field).
This also illustrates the intrinsic occurrence of coherence between the spectral components that built the quantum field of a Bose-Einstein condensate with 
relative phases between forward and backward propagating partial waves, as discussed in Ref.~\cite{ref-6}.
The interference of partial matter waves from the forward and backward propagating field components at the Boltzmann equilibrium intrinsically defines a time scale $\tau_0$, i.e. the propagation of the quantum field in units of this coherence time (or oscillation period in complex space),
the relative phase of the two counter-propagating wave fields does not appear to be completely random but coherent and quantized in phase around multiples of the circle number $2\pi$.

Consequently, coupling the two components at initial zero particle number imbalance with a Josephson weak link leads to the deterministic distribution of the relative initial phase and therefore to oscillations of the particle number imbalance.
The distribution of the relative initial phase around zero measure has been experimentally demonstrated in the framework of the experiments presented in Ref. \cite{ref-12}.
The latter aspect is experimentally confirmed by the distribution of phase correlations with quantization of the initial relative phase (unstable configurations of the relative phases) by measuring the 
Josephson oscillation period $\Omega$ which is defined in terms of the trap geometry and interaction strength on the one hand, as well as the distribution and the effect of the initial phase fluctuations on the energy splitting between the atomic ensembles \cite{ref-15}.

\section{Conclusion}

In conclusion, within the framework of a number-conserving quantum field theory with non-local order parameters, it is possible to derive quantitative estimates for the distribution of phase correlations
between two independently prepared Bose-Einstein condensates at thermal equilibrium ab initio. 
From the analysis of phase correlations, it is possible to illustrate that Josephson oscillations between two weakly coupled Bose-Einstein condensates occur due to the initial phase differences.
The presented model bears insights into the constructive and destructive interference of partial matter waves that lead to non-zero and variable averages of the spatially averaged quantum field in second quantization. 
From this definition of a non-local order parameter, it is possible to explain that the phase correlations between forward and backward-propagating components 
that span the quantum field of a single Bose-Einstein condensate at finite temperature is quantized in units of the circle number $2\pi$. 
The latter aspect arises from the thermalization of the atomic vapor to the Boltzmann equilibrium which is mathematically described as an internal measurement process (projection onto the Boltzmann equilibrium state) in the present model.
Numerical Monte Carlo simulations illustrate that this feature of the model builds the zero relative initial phase distribution.
Notably, the different frequency components can be experimentally measured and formally highlighted as a frequency comb spectrum for each realization of Josephson oscillation cycles with certain initial conditions from a composition 
of the oscillation frequency of the resulting Josephson current and the shift of the frequency spectrum from the initial phase imbalance. 

\acknowledgments

The author acknowledges the financial support from IU Internationale Hochschule for the (freelancer) lecturer position at the university, which has in particular enabled the formulation and editing of the present theory 
on Josephson oscillations of two weakly coupled Bose-Einstein condensates.

\end{document}